\documentclass[twocolumn,showpacs,prl]{revtex4}
\usepackage{graphicx}
\usepackage{dcolumn}
\usepackage{bm}
\usepackage{amsmath}
\usepackage{graphics}
\begin{document}

\title{Deterministic Weak Localization in Periodic Structures}

\author{C. Tian$^{1,2}$}
\author{A. Larkin$^{1,2,3}$}
\affiliation{$^{1}$ Kavli Institute for Theoretical Physics,
University of California, Santa Barbara, CA 93106, USA\\
$^{2}$ William I. Fine Theoretical Physics Institute, University
of Minnesota,
Minneapolis, MN 55455, USA\\
$^{3}$ L. D. Landau Institute for Theoretical Physics, Moscow,
117940, Russia}

\date{\today}

\begin{abstract}
The weak localization is found for perfect periodic structures
exhibiting deterministic classical diffusion. In particular, the
velocity autocorrelation function develops a universal quantum
power law decay at $4$ times Ehrenfest time, following the
classical stretched-exponential type decay. Such deterministic
weak localization is robust against weak enough randomness (e.g.,
quantum impurities). In the $1$D and $2$D cases, we argue that  at
the quantum limit states localized in the Bravis cell are turned
into Bloch states by quantum tunnelling.
\end{abstract}


\maketitle

The remarkable phenomenon of strong localization (SL) \cite{RL85}
nowadays has been fully appreciated in aperiodic \cite{RL85} and
periodic \cite{TA93} disordered electronic systems and disordered
dielectric materials \cite{lightloc}. The weak localization (WL),
as the precursor of SL \cite{AALR79} has been confirmed by
tremendous experiments in the last two decades, e.g., the negative
magnetoresistance \cite{RL85}, the coherent backscattering of
light \cite{WM85}, etc.. All these systems are strongly scattered,
i.e., $\tilde \lambda \gtrsim a$\,, where $\tilde \lambda$ is the
wavelength involved and $a$ is the typical scale over which the
potential varies substantially. Yet, in nineties unprecedented
degree of control reached in experiments with mesoscopic quantum
dots \cite{Alhassid}
allows to investigate the semiclassical region: $\tilde \lambda
\ll a$\,, where quantum interference is suppressed, and classical
chaotic motion prevails. Among central issues is the
classical-to-quantum crossover. A long time ago \cite{LOBZ} it was
established that appears the so-called Ehrenfest time marking the
proliferation of quantum interference. However, the quantitative
crossover had not been found until recently \cite{AL96}, in the
context of quantum corrections to the Drude conductivity of
disordered ballistic systems.

It is well known \cite{RL85,AL96} that in disordered time-reversal
symmetric systems, WL originates from the quantum interference
between two counter-propagating trajectories, and the diffusive
nature of trajectories arises from the randomness, regardless of
classical or quantum potentials. In contrast, the {\it
deterministic} diffusive motion may take place on a {\it classical
periodic} potential, where the extended Bloch state develops at
the quantum limit. Transport properties in such system remains
unexplored. On the other hand, the so-called periodic Lorentz gas,
as a prototype has been well understood mathematically
\cite{Sinai,Chernov,Gaspard96}, where the interaction with the
potential is simplified as ``mirror reflection''. Essential
properties are encapsulated in a series of theorems
\cite{Sinai,Chernov,Gaspard96}. To summarize, (i) the flow is $K$
mixing \cite{Sinai}; (ii) for piecewise H{\"o}lder continuous
(PHC) functions the correlation decay [e.g., the velocity
autocorrelation function (VCF)] is fast of stretched exponential
type, i.e., $\sim \exp (-t^\gamma)\,, 0<\gamma<1$ \cite{Chernov};
and (iii) the diffusive dispersion relation is identified as a
Pollicott-Ruelle resonance with an exact Green-Kubo relation for
the diffusion constant established \cite{Gaspard96}.
A finite diffusion constant is ensured by a priori, so-called
finite horizon condition \cite{Sinai}. To explore quantum
manifestations of these features naturally becomes a fundamental
problem.


The main finding of the present work is to show that such {\it
deterministic} classical diffusion leads to WL (to distinguish
from the usual WL in disordered systems, here we may term it {\it
deterministic WL}). In particular, at the one-loop level, we find
the frequency-dependent quantum correction to the diffusion
constant to be
\begin{equation} \delta D(\omega)
=-{D_{cl}\over {\pi \hbar \nu}}\, \Gamma(\omega)\!
 \int  \frac{d {\bf k}} {(2\pi)^2}\,
 \frac{1}{-i\omega+D_{cl} {\bf k}^2} \,,
                                                       \label{DDclKW}
\end{equation}
where $\nu$ is the density of states,
and the renormalization factor
\begin{equation}
\Gamma (\omega) =\exp \left(4i\omega t_E -\frac{4\omega ^2 \lambda
_2 t_E}{\lambda ^2}\right) \,. \label{Gamma1}
\end{equation}
The Ehrenfest time, say $t_E$ is
\begin{equation}
t_E=\lambda^{-1} \left|\ln \sqrt{\frac{a}{\lambda_F}}\right| \,.
\label{Ehren}
\end{equation}
Here $\lambda$ is the Lyapunov exponent and can be expressed in
the form of Abramov formula \cite{Gaspard96}. Depending on initial
conditions, $t_E$ may fluctuate, characterized by $\lambda_2$ with
the characteristic scale $\delta t_E =\lambda_2 t_E/ \lambda^2$\,.
Results similar to Eqs.~(\ref{DDclKW})-(\ref{Ehren}) were found
for disordered ballistic systems earlier \cite{AL96}. However, we
emphasize that throughout the derivation below no quantum
impurities will be introduced. Therefore, the main problem
surrounding the result of Ref.~\cite{AL96}, namely the possibility
of removing auxiliary quantum impurities is solved. On the other
hand, for the first time we see that at the quantum limit WL leads
to {\it extended } states rather than SL.

\begin{figure}[ht]
\begin{center}
\includegraphics[width=0.4
\textwidth]{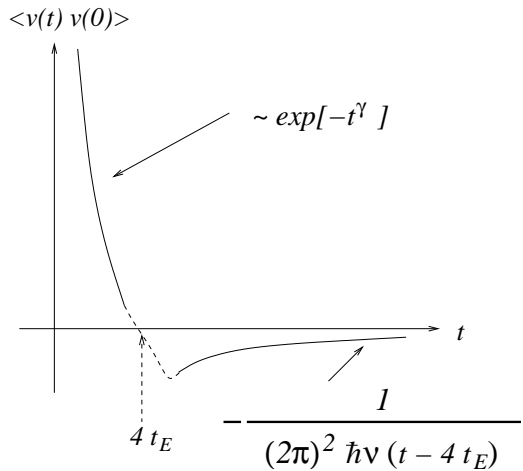}
\end{center}
\caption{The velocity autocorrelation function of the motion in a
$2$D periodic Lorentz gas (with finite horizon) develops a power
law decay at $t\gtrsim 4t_E$\,.} \label{WLorentzCorre}
\end{figure}

Alternatively, according to the Green-Kubo formula one may view
the result above from VCF $\langle {\bf v}(t) \cdot {\bf
v}(0)\rangle$ [with the average over the primitive cell of the
phase space energy shell (for shorthand we will call primitive
cell below.)]. In the semiclassical limit, $t_E$ satisfies
$\tau_{fl}\ll t_E \ll t_L$\,, where the localization time scale
$t_L \sim md\sqrt {D_{cl}/\hbar}$ and $\tau_{fl}$ is the {\it
classical} mean free time. Ignoring $\delta t_E$\,, we find that
the VCF develops a power law decay at $t \gtrsim 4 t_E$\ beyond
the {\it classical} stretched-exponential type decay.
Particularly, in the $2$D case, the power law takes the form as
$-(4 \pi^2\hbar\nu)^{-1} \, (t-4 t_E)^{-1}$ (see
Fig.~\ref{WLorentzCorre}).

Remarkably, the standard WL correction acquires a dispersion
$\Gamma (\omega)$\,, prevailing at $\omega \sim t_E^{-1}$\,. In
fact, similar renormalization factors exist at all the higher
order loop corrections. Such phenomenon originates from the
interference nature of the localization. Indeed, two classical
trajectories pass through (almost) the same position at (almost)
the same momentum. Subsequently they diverge and eventually take
counter-propagating routes. If these positions and momenta were
strictly identical, the probability would be zero. The quantum
uncertainty makes it possible. It takes, however, a long time
$\sim t_E$\,.

Having outline the qualitative picture, we turn to details of the
proof. To be specific, let us focus on the $2$D case. The model
consists of an electron of mass $m$ with the energy $E$ (namely
the Fermi wavelength $\lambda_F = \hbar/\sqrt{2mE}$) moving in a
periodic Lorentz gas. The lattice constant is $d$\,, while $a$ is
referred to the disk radius.
Technically, to avoid the singularity of the hard core potential,
the billiard is softened to be the potential $V({\bf r})$ around
the boundary with the typical width $b \ll W:= d-2a$\,. The
Hamiltonian thereby is replaced by ${\hat {\cal H}} := {\hat {\bf
p}}^2/2m + V(\bf r)$\,.

The basic tool that we will employ is the generalization of the
diagrammatical technique developed in essentially different
context--the kicked rotor \cite{TKL}. We will skip parallel
intermediate steps, with emphasis on the main difference. Let us
start from the {\it quantum} four-point ``density-density''
correlator, defined as: ${\cal D}({\bf r}_+,{\bf r}_-; {\bf
r}_+',{\bf r}_-';t,t') :=
  \langle {\bf r}_+| \exp[i {\hat {\cal H}} t/ \hbar]|{\bf r}_+' \rangle
\,  \langle {\bf r}_-|\exp[i {\hat {\cal H}} t'/ \hbar]|{\bf r}_-'
\rangle ^* $\,, where ${\bf r}_\pm\,, {\bf r}_\pm'$\,. A crucial
step is to introduce an artifitial ``one-step'' evolution operator
${\hat U} := \exp [i {\hat {\cal H}} t_* /\hbar]$\,, and consider
the density-density correlator at times of multiple $t_*$, i.e.,
$t=nt_*\,, t'=n't_*$\,. Here $t_*$ is a time scale such that $t_*
\ll \tau_{fl}$\,. Passing to the frequency
representation--$(\omega\,, E)$\,, the density-density correlator
is
\begin{eqnarray}
{\cal D}_E ({\bf r}_+,{\bf r}_-; {\bf r}_+',{\bf r}_-';\omega) &=&
\sum_{n,n'=0}^\infty \langle {\bf r}_+| {\hat U}^{n} |{\bf r}_+'
\rangle \, \langle {\bf r}_-|{\hat U}^{n'}
|{\bf r}_-' \rangle ^* \nonumber\\
&\times & e^{\frac{i\omega t_* (n+n')}{2}}\, e^{\frac{iE t_*
(n-n')}{\hbar}} \,. \label{QDiff}
\end{eqnarray}

\begin{figure}[ht]
\begin{center}
\includegraphics[width=0.47\textwidth]{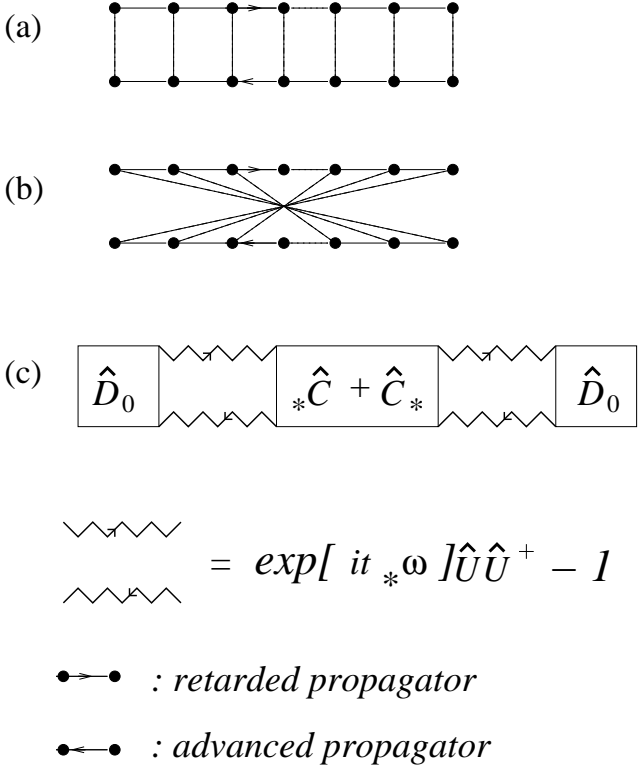}
\end{center}
\caption{Typical diagrams lead to the Diffuson (a) and the
Cooperon (b). The difference of two coordinates connected by the
dot-dashed line denotes is order of $\lambda_F \ll W$\,. The one
loop correction is given by (c). The retarded (advanced)
propagation line is  $\exp[i t_* \omega_+] {\hat U}$ ($\exp[-i t_*
\omega_-] {\hat U}^\dagger$). See Eq.~(\ref{CC}) for the
definition of $_* {\hat {\bf {\rm C}}}$ and ${\hat {\bf {\rm
C}}}_*$\,.} \label{WLorentzdiag}
\end{figure}

Now we consider the classical limit of the exact quantum
density-density correlator. First let us sum over all the diagrams
such that $|{\bf r}_+-{\bf r}_-|\,, |{\bf r}_+'-{\bf r}_-'|\,,
|{\bf r}_{k+}-{\bf r}_{k-}| \sim \lambda_F \ll W\,, k=1,2,\cdots$
[see Fig.~\ref{WLorentzdiag} (a)] and denote it as ${\cal D}_0$\,.
To proceed further, we perform Wigner transform the quantum
density-density correlator with respect to ${\bf r}_+-{\bf r}_-\,,
{\bf r}_+'-{\bf r}_-'$ and pass to the $({\bf p}\,, {\bf p}')$
representation. As a result, we find that ${\cal D}_0$ evolves
following the {\it classical} Perron-Frobenius equation according
to ($\omega$ will be ignored to shorten notations.)
\begin{eqnarray}
   &   &   \left\{ 1 - \exp \left[t_*\left(i\omega-{\hat {\cal L}}\right)\right] \right\} {\cal D}_{E0} ({\bf r}, {\bf p};
{\bf r}' ,{\bf p}')  \label{DiffClassical}\\
   &  =  &   \left( 2\pi \hbar\right)^{2} \delta \left({\bf r}-{\bf
r}'\right) \delta \left({\bf p}-{\bf p}'\right) \delta \left(E
-\frac{{\bf p}'^2}{2m}\right)\, , \nonumber
\end{eqnarray}
where ${\hat {\cal L}}= {\bf v}\cdot \nabla_{\bf r}-\nabla_{\bf r}
V({\bf r})\cdot \nabla_{\bf p}$ is the Liouvillian. Note that in
most region of the phase space the Liouvillian is free except near
the boundary of the billiard, where the potential is involved.
According to Eq.~(\ref{DiffClassical}), the modified Hamiltonian
induces a flow restricted on the energy shell: $E={\bf p}'^2
/2m$\,. We assume that for this flow, the essential properties
(i), (ii) and (iii) for the periodic Lorentz gas
hold. To be specific, {\it in the PHC functional space, the
Fourier transform of the operator $\{ 1 - \exp [t_*(i\omega-{\hat
{\cal L}})]\}^{-1}$ admits the diffusive poles: $\omega=-iD_{cl}
{\bf k}^2$ (${\bf k}$ the wave number taken from the first
Brillouin zone of the reciprocal lattice). Moreover, the classical
diffusion constant, $D_{cl}$ is given by the Green-Kubo formula,
i.e.,
\begin{equation}
D_{cl}=\frac{1}{4}\, \frac{t_*}{m^2}\, \sum_{j=-\infty}^\infty \,
\left\langle {\bf p}(j)\cdot {\bf p}(0) \right\rangle
\label{GreenKubo}
\end{equation}
with the average over the primitive cell.} We point out that
$D_{cl}$ is finite due to the nature of the finite horizon,
although we leave the problem of calculating Eq.~(\ref{GreenKubo})
open. Based on this conjecture, despite of the deterministic
nature of Eq.~(\ref{DiffClassical}), one may work on distributions
with the smoothness admitted by the PHC condition. Define ${\cal
D}_{E0}({\bf r}, {\bf p}; {\bf r}', {\bf p}') := {\cal D}_0({\bf
r}, {\bf n}; {\bf r}', {\bf n}')\, \delta (E-{\bf p}^2 /(2m))\,
\delta (E-{\bf p}'^2 /(2m))$\,. Averaging ${\cal D}_0$ over the
primitive cell, and passing to the Fourier representation,
eventually we reduce ${\cal D}_0$ into the diffusive propagator
(denoted as ${\cal D}_\nu$)
\begin{eqnarray}
{\cal D}_\nu \left({\bf k};\omega\right) = \frac{2\pi}{\nu}\,
\frac{1}{-i t_* \omega + t_* D_{cl} {\bf k}^2} \label{DiffuApp}
\end{eqnarray}

We note that in fact, to sum over all the diagrams such that
$|{\bf r}_+-{\bf r}_-'|\,, |{\bf r}_+'-{\bf r}_-|\,, |{\bf
r}_{k+}-{\bf r}_{(n+1-k)-}| \sim \lambda_F \ll W\,,
k=1,2,\cdots,n$ [Fig.~\ref{WLorentzdiag} (b)] yields another
classical object, so-called Cooperon (denoted as ${\cal C}_0$). In
fact, one may Wigner transform the quantum density-density
correlator with respect to ${\bf r}_+-{\bf r}_-'\,, {\bf
r}_+'-{\bf r}_-$ and arrive at Eq.~(\ref{DiffClassical}).

Having identified the classical {\it normal} diffusive modes, we
come to analyze WL. For this purpose, the essential step is to
show that in the semiclassics, the one-loop quantum correction is
given by the diagram shown in Fig.~\ref{WLorentzdiag} (c). In the
exact quantum operator representation, it may be expressed as
\begin{equation}
\delta {\hat {\cal D}}= {\hat {\cal D}}_{E0} \, \left({\hat {\cal
P}} -1 \right) \, \left(_{*} {\hat { {\bf C}}}+{\hat { {\bf C}}}_*
\right)\, \left({\hat {\cal P}} -1 \right) \, {\hat {\cal D}}_{E0}
\label{oneloop}
\end{equation}
with
\begin{equation}
_{*} {\hat { {\bf C}}} := \sum_{n=1}^\infty  {\hat {\cal P}} \,
^{n+1} \, {\hat {\cal C}}_{E0} \, {\hat {\cal P}} \,, \quad {\hat
{ {\bf C}}}_{*} := \sum_{n=1}^\infty {\hat {\cal P}}  \, {\hat
{\cal C}}_{E0} \, {\hat {\cal P}} \, ^n \,. \label{CC}
\end{equation}
Here ${\hat {\cal P}}= e^{i\omega t_*} {\hat U}{\hat
U}^\dagger$\,. Notice that in Eq.~(\ref{oneloop}) the operator
${\hat {\cal P}}$ in $({\hat {\cal P}}-1)$ is chosen in the way
such that it describes the free motion. As usual, we then perform
the Wigner transform for ${\hat {\cal D}}_{E0}$\,, $_* {\hat {\bf
C}}$\,, ${\hat {\bf C}}_*$ and ${\hat {\cal P}}$, respectively.
Now we assume that the initial distribution is PHC. Furthermore,
the appearance of diffusive poles allows us to make the so-called
hydrodynamic expansion, $\omega t_* \ll \omega \tau_{fl} \ll 1$
and $\exp [t_* (i\omega - {\hat {\cal L}})]-1 \approx -t_* {\bf
v}\cdot \nabla$\,. Eventually in the PHC functional space, the
one-loop quantum correction to the Diffuson is reduced into
\begin{eqnarray}
\delta {\cal D}\left({\bf r},{\bf n};{\bf r}',{\bf n}'\right)=
  \frac{1}{2}\,\frac{t_*^2 E}{m} {\hat {\bf { V}}} \, \bigg [ \sum_{j=-\infty}^\infty \,
{\bf n}''(j)\, {\bf n}''(0)
\nonumber\\
  :  \nabla_{{\bf r}_1}\, \nabla_{{\bf r}_2} {\cal D}_0\left({\bf r},{\bf n};{\bf r}_1+\frac{\delta {\bf
r}_2}{2},{\bf n}''+\frac{\delta {\bf n}_1}{2}\right) \nonumber\\
\times {\cal D}_0\left({\bf r}_2-\frac{\delta {\bf r}_2}{2},-{\bf
n}''+\frac{\delta{\bf n}_1}{2};{\bf r}',{\bf n}'\right)\bigg
]\bigg |_{{\bf r}_1={\bf r}_2={\bf r}''}\, , \label{exactvertex4}
\end{eqnarray}
where ${\bf n}(0):={\bf n}''$\,. The vertex operator ${\hat {{\bf
{ V}}}}$ is an integral operator
defined as
\begin{eqnarray}
 &   &   \left({\hat {\bf { V}}} f \right)\, (\cdot)
     :=    \int  \frac{d{\bf r}''d{\bf n}''}{2 \pi \hbar}
\int \frac{d\delta {\bf r}_1d\delta{\bf n}_1}{2 \pi \hbar} \int
\frac{d\delta {\bf r}_2d\delta{\bf n}_2}{2 \pi \hbar}
\nonumber\\
 & \times &  {\cal C}_0\left({\bf r}''+\frac{\delta
{\bf r}_1}{2},{\bf n}''-\frac{\delta{\bf n}_2}{2};{\bf
r}''-\frac{\delta {\bf r}_1}{2},-{\bf n}''-\frac{\delta{\bf
n}_2}{2}\right) {\cal X}
\nonumber\\
 & \times &   f\, \left(\cdot \,; {\bf r}'', {\bf n}''\,
;\delta {\bf r}_2, \delta {\bf n}_1 \right) \,,  \label{interactionV}
\end{eqnarray}
where
\begin{equation}
{\cal X} = \exp \left[\frac{i}{\lambda_F} \left( \delta {\bf r}_1
\cdot \delta{\bf n}_1 + \delta {\bf r}_2 \cdot \delta{\bf n}_2
\right)\right] \,. \label{X}
\end{equation}
Above the dot is the shorthand of $({\bf r},{\bf n};{\bf r}',{\bf
n}')$\,. The interference factor ${\cal X}$ originates from the
phase difference of two paths underlying the Cooperon, and
technically results from the difference of performing Wigner
transform for the Diffuson and the Cooperon. ${\cal X}$ can be
also obtained in the Moyal formalism \cite{MA04}. In the diffusive
limit, Eq.~(\ref{exactvertex4}) leads to the standard WL in
disordered systems. For ballistic systems like periodic Lorentz
gases,
the two legs of the Cooperon must propagate together for $2t_E$ in
the Lyapunov region in order to develop the universal quantum
limit of Cooperon develops, which stands for the probability of
returning to the origin via the random walk. Such a crossover
picture, encapsulated in the renormalization factor
$\Gamma(\omega)$ was proposed earlier \cite{AL96}. Here the
essential difference is that, due to the absence of quantum
impurities the motion equation of Cooperon does not involve any
regularizers. Consequently the Ehrenfest time is fully determined
by the size of the minimal quantum wave packet, as read out from
${\cal X}$\,, i.e., $\delta {\bf r}_{1,{\rm min}}\sim \lambda_F$
and $\delta {\bf n}_{2,{\rm min}}\sim \lambda_F/W$\,. Then
following the procedure of Ref.~\cite{TKL} one ends up with
Eqs.~(\ref{Gamma1}) and (\ref{Ehren}).

At the time $\sim t_E$\,, $f$ acted by ${\hat {\bf V}}$ is smeared
out over the primitive cell surrounding the entries ($f$ thereby
obtained is denoted as $f_\nu$). Consequently it can be shown that
the interaction vertex ${\hat {\bf V}}$ can be simplified as
(denoted as ${\hat {\bf V}}_\nu$)
\begin{eqnarray}
\left({\hat {\bf V}} f \right)\, \left(\cdot\right)
   &  \rightarrow  &
\left({\hat {\bf V}}_\nu f_\nu \right)\, \left(\cdot\right)     :=
{\bf V} \int d{\bf r}'' \, f_\nu \left({\bf r}, {\bf r}'; {\bf
r}''\right) \,,
\nonumber\\
{{\bf { V}}}   &   =   &   \frac{\Gamma(\omega)}{\pi \hbar \nu} \!
 \int  \frac{d {\bf k}}{(2\pi)^2}\,
 \frac{1}{-i\omega+D_{cl}{\bf k}^2} \,.
                                                       \label{VRenormalization}
\end{eqnarray}
Now let us identify the Diffusons in the right hand side of
Eq.~(\ref{exactvertex4}) as {\it deterministic} diffusive modes.
Technically, we may average $({\bf r}, {\bf n})$ and $({\bf r}',
{\bf n}')$ over the primitive cell surrounding them. With the
application of Eq.~(\ref{VRenormalization}), we thereby obtain the
one-loop quantum correction as
\begin{eqnarray}
\delta {\cal D}\left({\bf r}, {\bf r}'\right)
  &  =  &   t_* D_{cl}\, {{\bf {V}}}
\int d{\bf r}''\, \left(\nabla_{{\bf r}_1}^2+\nabla_{{\bf
r}_2}^2\right)
\nonumber\\
  &     &    \left[{\cal D}_\nu \left({\bf r}, {\bf r}_1\right) {\cal
D}_\nu \left({\bf r}_2, {\bf r}'\right)\right] |_{{\bf r}_1={\bf
r}_2={\bf r}''}\,, \label{oneloopsimplification}
\end{eqnarray}
where we take into account the isotropy of the Bravis lattice.
Then we perform the Fourier transform and substitute
Eq.~(\ref{DiffuApp}), as well as the renormalized interaction
vertex ${\bf V}$ [see Eq.~(\ref{VRenormalization})] into it. As a
result, we find that the diffusive modes are modified as $[-i t_*
\omega  + t_* (D_{cl}+\delta D(\omega)) {\bf k}^2]^{-1}$ with
$\delta D(\omega)$ given by Eq.~(\ref{DDclKW}).

An important question is how WL develops into {\it extended}
states in the quantum limit. For qualitative discussions, it may
be instructive to discuss the quasi-$1d$ case. 
For $\lambda_F \ll W \ll d$\,, each Bravis cell has a well defined
mean level spacing, say $\Delta \sim \hbar^2 /m d^2 $\,, with a
tunnelling energy, say $t$ between them. Increasing the size of
the cell $n$ times larger then we rescales the mean level spacing
as $\Delta_n \sim \Delta/n $\,. For $t_n$\,, we note that the
Thouless energy of the rescaled cell is of the order of the total
tunnelling energy, namely $\hbar D_{cl}/ (nd)^2 \sim
t_n^2/\Delta_n $\,, yielding $t_n \sim ({\hbar D_{cl}
\Delta_n})^{1/2} / nd $\,. Localization states develop at
$\Delta_n \gtrsim t_n$\,, namely $n\gtrsim n_c$ with $n_c \sim m
D_{cl}/\hbar \sim W/\lambda_F$ the localization length. In
contrast to the usual disordered system, these localization states
(with the same energy) are exactly {\it identical} because of the
periodic feature, and thus, serves as Wannier function in the
electronic band theory. The tunnelling between these states turns
them into the extended state. In particular, at the quantum limit,
$W\sim \lambda_F $\,, $n_c\sim 1$\,, the energy band takes the
form of Bloch band, i.e., $\varepsilon + t_{n_c} \cos (d k/2\pi)
$\,.

In realistic environments randomness (e.g., quantum impurities)
dooms to exist. Bloch states are unstable against randomness.
Indeed, in the $1$D and $2$D cases arbitrarily weak randomness
leads to SL at the long time (or large size) limit. In contrast,
the deterministic WL is robust. In fact, any randomness results in
a scattering rate $\tau_q^{-1}$\,. As far as it is weak enough,
namely $\tau_q^{-1} \ll (\hbar \nu)^{-1}$\,, the Ehrenfest time is
dominated by the size of the initial quantum wavepacket due to the
logarithmic accuracy. This conclusion, indeed is consistent with
an important conjecture made in Ref.~\cite{AL96}. That is,
although auxiliary quantum impurities may be introduced
artificially, the strength of the scattering rate $\tau_q^{-1}$
must be adjusted to be $\sim (\hbar \nu)^{-1}$ in order to mimic
quantum diffractions. Thus, an estimation for the Ehrenfest time
was established \cite{AL96}, which is the same as
Eq.~(\ref{Ehren}). In the present work, this essential conjecture
is proved.

To conclude, we find WL for periodic Lorentz gases with finite
horizon. The result is applicable for any systems exhibiting
deterministic diffusive motion, regardless whether the underlying
potential is ordered or disordered. Finally since WL is
responsible for coherent backscattering \cite{AWM86}, we expect
that the result here may be confirmed by experiments on coherent
backscattering of light in state-of-art periodic photonic
materials, 
where the semiclassical condition: $\tilde \lambda \ll a$ may be
met.

We are grateful to A. Kamenev for numerous important discussions
and A. Altland, J. Cao, K. B. Efetov, F. Haake, J. M{\"u}ler and
Ya. G. Sinai for useful conversations. This work is supported by
NSF under Grants No. DMR-0120702, DMR-0405212 and PHY-9907949.

\end{document}